\documentclass[a4paper,noarxiv,amsmath,twocolumn]{quantumarticle}
\usepackage[ascii]{inputenc}
\usepackage{amsmath,amssymb,amsfonts,amsthm}
\usepackage{mathtools}
\usepackage[english]{babel}
\usepackage{braket}
\usepackage{graphicx}
\usepackage[caption=false]{subfig}
\usepackage[numbers,compress]{natbib}
\usepackage{algorithm}
\usepackage{algorithmicx}
\usepackage{algpseudocode}
\usepackage{todonotes}
\usepackage{tikz}
\usepackage{float}
\usepackage{placeins}
\usetikzlibrary{quantikz}
\usepackage[pdftex,bookmarks=false]{hyperref}
\usepackage{tikz-cd}
\usepackage{footnote}
\tikzcdset{scale cd/.style={every label/.append style={scale=#1},
    cells={nodes={scale=#1}}}}

\DeclarePairedDelimiter\abs{\lvert}{\rvert}
\DeclarePairedDelimiter\norm{\lVert}{\rVert}

\DeclareMathOperator{\Tr}{Tr}

\theoremstyle{definition}
\newtheorem{definition}{Definition}[section]

\begin{document}
\title{Quantum Wasserstein GANs for State Preparation at Unseen Points of a Phase Diagram}

\author{Wiktor Jurasz}
\affiliation{Technical University of Munich, CIT, Department of Computer
  Science, Boltzmannstra{\ss}e 3, 85748 Garching, Germany}
\thanks{Now at Google.}

\author{Christian B.~Mendl}
\affiliation{Technical University of Munich, CIT, Department of Computer Science, Boltzmannstra{\ss}e 3, 85748 Garching, Germany}
\affiliation{Technical University of Munich, Institute for Advanced Study, Lichtenbergstra{\ss}e 2a, 85748 Garching, Germany}

\begin{abstract}
Generative models and in particular Generative Adversarial Networks (GANs) have
become very popular and powerful data generation tool. In recent years, major
progress has been made in extending this concept into the quantum realm.
However, most of the current methods focus on generating classes of states that were
supplied in the input set and seen at the training time.
In this work, we propose a new hybrid classical-quantum method based on quantum
Wasserstein GANs that overcomes this limitation. It allows to learn the function
governing the measurement expectations of the supplied states and generate new
states, that were not a part of the input set, but which expectations follow the
same underlying function.
\end{abstract}

\section{Introduction}

Classical Generative Adversarial Networks (GANs) are formulated as min-max game between a ``generator'' and a ``discriminator''. They exhibit surprising capabilities, like the generation of photorealistic pictures of non-existing persons \cite{goodfellow2014}.
The mathematical framework and practical training procedure of these networks has been improved using the Kantorovich dual formulation from optimal transport as distance metric, leading to so-called Wasserstein GANs \cite{arjovsky2017, gulrajani2017}.

An intriguing research direction is the translation of these concepts to the quantum realm.
Initiated by pioneering work \cite{lloyd2018, dallaire2018}, this goal continues to be pursued from several angles \cite{benedetti2019, zeng2019, lu2020, herr2021, ahmed2021}.
Regarding Wasserstein GANs in particular, the recently proposed and analyzed quantum Wasserstein distance \cite{depalma2021} ameliorates the issue of vanishing gradients affecting other, unitarily invariant distance metrics, cf.~\cite{chakrabarti2019, Becker2021}.
A central concept in \cite{depalma2021} are ``neighboring'' quantum states, defined as density matrices which agree after discarding (tracing out) a single qudit.
Kiani et al.~\cite{kiani2021} have demonstrated the practical advantages and feasibility of corresponding quantum Wasserstein GANs (qWGANs) for ``learning'' a target quantum state.

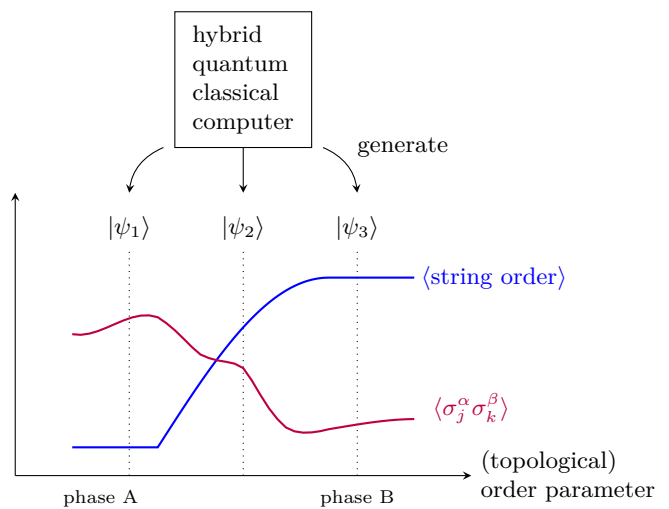
\begin{figure}[!ht]
\centering
\begin{tikzpicture}[>=stealth, scale=1.5]
\draw[->] (0, 0) -- (4, 0) node[right, align=left, font=\small] {(topological) \\ order parameter};
\draw[->] (0, 0) -- (0, 2.5);
\node[font=\scriptsize] at (0.75, -0.2) {phase A};
\node[font=\scriptsize] at (3,    -0.2) {phase B};
\draw (1.4, 2.9) rectangle (2.6, 4.1);
\node[align=left, font=\small] at (2, 3.5) {hybrid \\ quantum \\ classical \\ computer};
\node[font=\small] at (3.4, 2.9) {generate};
\draw[->] (1.3, 2.9) to[bend right] (1, 2.5);
\draw[->] (2,   2.9) to (2, 2.5);
\draw[->] (2.7, 2.9) to[bend left] (3, 2.5);
\draw[dotted] (1, 0) -- (1, 2) node[above, font=\small] {$\ket{\psi_1}$};
\draw[dotted] (2, 0) -- (2, 2) node[above, font=\small] {$\ket{\psi_2}$};
\draw[dotted] (3, 0) -- (3, 2) node[above, font=\small] {$\ket{\psi_3}$};
\draw[blue, thick] (0.5, 0.25) -- (0.5375, 0.25) -- (0.575, 0.25) -- (0.6125, 0.25) -- (0.65, 0.25) -- (0.6875, 0.25) -- (0.725, 0.25) -- (0.7625, 0.25) -- (0.8, 0.25) -- (0.8375, 0.25) -- (0.875, 0.25) -- (0.9125, 0.25) -- (0.95, 0.25) -- (0.9875, 0.25) -- (1.025, 0.25) -- (1.0625, 0.25) -- (1.1, 0.25) -- (1.1375, 0.25) -- (1.175, 0.25) -- (1.2125, 0.25) -- (1.25, 0.25) -- (1.2875, 0.30889) -- (1.325, 0.367689) -- (1.3625, 0.426306) -- (1.4, 0.484652) -- (1.4375, 0.542635) -- (1.475, 0.600168) -- (1.5125, 0.657161) -- (1.55, 0.713525) -- (1.5875,  0.769176) -- (1.625, 0.824025) -- (1.6625, 0.87799) -- (1.7, 0.930986) -- (1.7375, 0.982932) -- (1.775, 1.03375) -- (1.8125, 1.08336) -- (1.85, 1.13168) -- (1.8875, 1.17864) -- (1.925, 1.22417) -- (1.9625, 1.2682) -- (2., 1.31066) -- (2.0375, 1.35148) -- (2.075, 1.39061) -- (2.1125, 1.42798) -- (2.15, 1.46353) -- (2.1875, 1.4972) -- (2.225, 1.52896) -- (2.2625, 1.55874) -- (2.3, 1.58651) -- (2.3375, 1.61221) -- (2.375, 1.63582) -- (2.4125, 1.65729) -- (2.45, 1.67658) -- (2.4875, 1.69368) -- (2.525, 1.70855) -- (2.5625, 1.72118) -- (2.6, 1.73153) -- (2.6375, 1.7396) -- (2.675, 1.74538) -- (2.7125, 1.74884) -- (2.75, 1.75) -- (2.7875, 1.75) -- (2.825, 1.75) -- (2.8625, 1.75) -- (2.9, 1.75) -- (2.9375, 1.75) -- (2.975, 1.75) -- (3.0125, 1.75) -- (3.05, 1.75) -- (3.0875, 1.75) -- (3.125, 1.75) -- (3.1625, 1.75) -- (3.2, 1.75) -- (3.2375, 1.75) -- (3.275, 1.75) -- (3.3125,  1.75) -- (3.35, 1.75) -- (3.3875, 1.75) -- (3.425, 1.75) -- (3.4625, 1.75) -- (3.5, 1.75);
\node[font=\small] at (4.2, 1.75) {\color{blue} $\langle\text{string order}\rangle$};
\draw[purple, thick] (0.5, 1.25) -- (0.575, 1.24448) -- (0.65, 1.25504) -- (0.725, 1.27736) -- (0.8, 1.30712) -- (0.875, 1.34) -- (0.95, 1.37168) -- (1.025, 1.39784) -- (1.1, 1.41416) -- (1.175, 1.41632) -- (1.25, 1.4) -- (1.325, 1.34792) -- (1.4, 1.28456) -- (1.475, 1.2116) -- (1.55, 1.136) -- (1.625, 1.07) -- (1.7, 1.03832) -- (1.775, 1.01816) -- (1.85, 1.00568) -- (1.925, 0.98936) -- (2., 0.95) -- (2.075, 0.84116) -- (2.15, 0.71408) -- (2.225, 0.59564) -- (2.3, 0.49928) -- (2.375, 0.425) -- (2.45, 0.39128) -- (2.525, 0.37964) -- (2.6, 0.3818) -- (2.675, 0.3926) -- (2.75, 0.41) -- (2.825, 0.422672) -- (2.9, 0.435296) -- (2.975, 0.447584) -- (3.05, 0.459248) -- (3.125, 0.47) -- (3.2, 0.479552) -- (3.275, 0.487616) -- (3.35, 0.493904) -- (3.425, 0.498128) -- (3.5, 0.5);
\node[font=\small] at (4, 0.6) {\color{purple} $\langle \sigma_j^{\alpha} \sigma_k^{\beta}\rangle$};
\end{tikzpicture}
\caption{Conceptual overview of the method proposed in this work: a collection of observables, including correlation functions and topological order indicators, are used in the dual formulation of the quantum Wasserstein distance by a hybrid quantum-classical computer to generate (unseen) quantum states. This allows, for example, to ``tune'' through a (topological) phase transition.}
\label{fig:qGAN_interpolation}
\end{figure}

Here we take these ideas a step further and develop a method for generating quantum states at hitherto unseen (topological) phase diagram points on a hybrid quantum-classical computer, as conceptually illustrated in Fig.~\ref{fig:qGAN_interpolation}.
Our method employs selected observables (serving as phase indicators) for the dual formulation of qWGANS.
We assume that their expectation values are at least continuous functions with respect to phase diagram parameters.

\section{Quantum Wasserstein distance}

As in the classical case, the qWGANs rely on calculating the Wasserstein
distance between the target and the generated state. There have been multiple
approaches to defining the Wasserstein distance in the quantum realm \cite{carlen2014,
 ning2015, chen2018a, chen2018b, golse2021, yu2019, depalma2021}.
In this work, we use the definition by De Palma et al.~\cite{depalma2021}. They
proposed a generalization of the Wasserstein distance of order 1 based on
``neighboring'' quantum states. 
The general setting is a quantum system of $n$ qudits, with corresponding Hilbert space $\mathcal{H}_n = (\mathbb{C}^d)^{\otimes n}$. Following the notation in \cite{depalma2021}, $\mathcal{O}_n$ denotes the set of Hermitian matrices acting on $\mathcal{H}_n$, $\mathcal{O}_n^+ \subset \mathcal{O}_n$ the subset of positive semidefinite matrices, and $\mathcal{S}_n \subset \mathcal{O}_n^+$ the set of density matrices (i.e., with unit trace).

\begin{definition}[Neighboring quantum states \cite{depalma2021}]
Two quantum states $\rho$ and $\sigma$ in $\mathcal{S}_n$ are \emph{neighboring} states if they coincide after discarding one qudit, i.e., $\Tr_i[\rho] = \Tr_i[\sigma]$ for some $i \in \{ 1, \dots, n \}$, where $\Tr_i$ denotes the partial trace over the $i$-th qudit.
\end{definition}
Informally, the $W_1$ distance is the maximum distance that is induced by a norm that assigns the distance at most one to any couple of neighboring states.

Formally, the $W_1$ distance is defined as \cite{depalma2021}:
\begin{equation}
\begin{split}
&W_1(\rho, \sigma) \\
&= \min\Bigg(\sum_{i=1}^n c_i : c_i \geq 0, \rho - \sigma = \sum_{i=1}^n c_i\left(\rho^{(i)} - \sigma^{(i)}\right), \\
&\hspace{1.75cm} \rho^{(i)}, \sigma^{(i)} \in \mathcal{S}_n, c_i \in \mathbb{R}, \Tr_i\rho^{(i)} = \Tr_i\sigma^{(i)}\Bigg).
\end{split}
\end{equation}
Expressed by the corresponding dual formulation,
\begin{equation}
W_1(\rho, \sigma) = \max_{H \in \mathcal{O}_n} \big\{ \Tr[(\rho - \sigma)H] : \norm{H}_L \leq 1 \big\},
\label{eq:w1_distance_general}
\end{equation}
where $\norm{H}_L$ is the quantum Lipschitz constant of the matrix $H$, defined as:
\begin{equation}
\norm{H}_L = 2 \max_{i=1,\ldots,n} \min_{H_{\bar{i} \in \mathcal{O}_n}} \norm{H - H_{\bar{i}}}_\infty.
\label{eq:quantum_lipschitz_constant}
\end{equation}
Here $H_{\bar{i}}$ is a Hermitian matrix that does not act on the $i$-th qudit.
The quantum Lipschitz constant and the Wasserstein distance defined in this way recover their classical counterparts for operators diagonal in the canonical basis.

The quantum Wasserstein distance has several properties that make it particularly useful in the context of training generative models:
\begin{enumerate}
\item It is invariant with respect to qudit permutations and super additive with respect to tensor products, i.e., $W_1(\rho, \sigma) \geq W_1(\rho_{1 \ldots m}, \sigma_{1 \ldots m}) + W_1(\rho_{(m+1) \ldots n},\sigma_{(m+1) \ldots n})$.
Here $\rho_{1 \ldots m}$ denotes a quantum state made out of qudits from index
$1$ to $m$, $W_1(\rho_{1 \ldots m}, \sigma_{1 \ldots m})$ denotes the
Wasserstein distance between those marginal states, and $n$ is a total number of
qudits in states $\rho$ and $\sigma$.
This property implies that an operation which reduces the distance between some marginal states also reduces the distance between the full states, for example, $W_1(\ket{100}, \ket{111}) > W_1(\ket{110}, \ket{111})$. Note that the fidelity does not have this property since it is always zero for orthogonal states.
\item The quantum Wasserstein distance is bounded by the trace distance, i.e., $\frac{1}{2} \norm{\rho - \sigma}_1 \leq W_1(\rho, \sigma) \leq \frac{n}{2} \norm{\rho - \sigma}_1$, where $n$ is the number of qudits. This ensures that minimizing the $W_1$ distance also minimizes the trace distance.
\item Because the quantum Wasserstein distance recovers the classical Wasserstein distance for diagonal operators, we can expect that generative models built using this metrics preserve the advantages of their classical counterparts.
\end{enumerate}

\section{Base algorithm and numerical method}

\subsection{qWGAN architecture}

Directly using Quantum Wasserstein distance to implement qWGANs is infeasible
because of the size of $\mathcal{O}_n$ in Eq.~\eqref{eq:w1_distance_general} and
Eq.~\eqref{eq:quantum_lipschitz_constant}.
In this section we layout in details a practical qWGANs
algorithm proposed by Kiani et al.~\cite{kiani2021}. We describe the
discriminator and generator architecture and how those two are trained.

This method allows to generate quantum states seen at training time. In the
next section we extend the algorithm to generate new, unseen states.

\subsubsection{Discriminator}
The discriminator architecture directly follows from
Eq.~\eqref{eq:w1_distance_general} and takes the form of a simple linear
program. In practice, one has to restrict the set $\mathcal{O}_n$ in
Eq.~\eqref{eq:w1_distance_general} to make the computation feasible.
As proposed in \cite{kiani2021}, the set of parametrized Pauli strings of length
$k$ is used. Specifically, let
\begin{equation}
\begin{split}
\label{eq:parametrized_hamiltonian}
H(W) &= \sum_{i_1 = 1}^{n-k+1} \ldots \sum_{i_k = i_{k-1} + 1}^{n} \\
&\quad \times \sum_{\sigma_1, \ldots, \sigma_k \in \{I, X, Y, Z\}} w^{\sigma_1, \ldots, \sigma_k}_{(i_1, \ldots, i_k)} \sigma^1_{i_1} \otimes \ldots \otimes \sigma^k_{i_k} \\
&= \sum_{\mathcal{I}_k \subseteq \{1, \ldots, n \}} \sum_{H_{\mathcal{I}_k} \in \{ I,X,Y,Z \}^{\otimes k}} w_{\mathcal{I}_k}^H H_{\mathcal{I}_k},
\end{split}
\end{equation}
where $\mathcal{I}_k$ is a $k$-set of qudit indexes used to generate the
length-$k$ Pauli string, $H_{\mathcal{I}_k}$ is the length-$n$ Pauli string that
acts non trivially on the set of at most $k$ qudits corresponding to
$\mathcal{I}_k$ and $W$ is the set of all weights.
To simplify the notation, the parameter set $W$ is enumerated as $W = \{w_1,
\ldots, w_N\}$, where $N = \abs{W}$ and $H_i$, $\mathcal{I}_{k_i}$ are the
Hamiltonian and index set associated with the weights $w_i \in \mathbb{R}$.

Now, the quantum Lipschitz constant in Eq.~\eqref{eq:quantum_lipschitz_constant} of $H(W)$ is bounded by
\begin{multline}
\label{eq:quantum_lipschitz_bound}
\norm{H(W)}_L \le \\
2 \max_{i = 1,\ldots,n} \norm*{\sum_{\substack{ \mathcal{I}_k \subseteq \{1, \ldots, n \} \\ \land i \in \mathcal{I}_k }} \sum_{\substack{H \in \{ I,X,Y,Z \}^{\otimes k} \\ H_{i} \neq I}} w_{\mathcal{I}_k}^H H_{\mathcal{I}_k} }_{\infty}
\end{multline}
where the sum is taken only over the operators which act non-trivially on qudit $i$ \cite{kiani2021}.

Eq.~\eqref{eq:w1_distance_general} now can be rewritten as in
Eq.~\eqref{eq:quantum_wasserstein_parametrized}, where the optimization is
performed with respect to the weights $w$ instead of the Hamiltonian set
$\mathcal{O}_n$:
\begin{equation}
\label{eq:quantum_wasserstein_parametrized}
W_1(\rho, \sigma) = \max_{w} \Tr\bigg[(\rho - \sigma)\sum_{i=1}^N w_i H_i\bigg],
\end{equation}
under the constraint stemming from the quantum Lipschitz constant bound in Eq.~\eqref{eq:quantum_lipschitz_bound}:
\begin{equation}
\label{eq:absolute_value_constraint}
\sum_{i \in \{1,\ldots, n\} \land j \in \mathcal{I}_{k_i}} \abs{w_i} \leq 1, \qquad j = 1,\ldots,n.
\end{equation}
This optimization problem can be translated into the canonical form of linear
programming. First, let
\begin{equation}
c_i = \Tr[(\rho - \sigma) H_i],
\end{equation}  
then Eq.~\eqref{eq:quantum_wasserstein_parametrized} becomes
\begin{equation}
W_1(\rho, \sigma) = \max_{w} \sum_{i=1}^N w_i c_i.
\end{equation}
Together with
\begin{equation}
w_i = w_i^+ - w_i^-,
\end{equation}
the absolute value constraint from Eq.~\eqref{eq:absolute_value_constraint} is equivalent to the following set of
constraints:
\begin{subequations}
\begin{align}
w_i^+ &\ge 0 \\
w_i^- &\ge 0 \\
\sum_{i \in \{1,\ldots, n\} \land j \in \mathcal{I}_{k_i}} \left(w_i^+ + w_i^-\right) &\leq 1, \qquad j = 1, \ldots, n
\end{align}
\end{subequations}
Now, with the two vectors defined as:
\begin{align}
w' &= [w_1^+, w_1^-, \ldots, w_N^+, w_N^-] \\
c' &= [c_1, -c_1, \ldots, c_N, -c_N],
\end{align}
and the matrix $A^{n \times N}$ defined as:
\begin{equation}
A_{j, i}  = \begin{cases}
    1 & \text{if} \quad j \in \mathcal{I}_{k_i} \\
    0 & \text{else}
\end{cases}
\end{equation}
the linear program for the discriminator in the canonical form reads:
\begin{equation}
\label{eq:linear_program_canonical}
\begin{split}
\max_{w'} \quad & c'^T w'\\
\text{subject to } \quad & w' \geq 0 \\
& A w' \leq 1 \quad \text{(pointwise)}.
\end{split}
\end{equation}

The weights from the original set $W$ are recovered as:
\begin{equation}
  w_i = w_{2i-1}' - w_{2i}'.
\end{equation}

The linear program with $n$ constraints outputs at most $n$ non-zero weights \cite{bertsimas1997}, so the optimal Hamiltonian which approximates the quantum Wasserstein distance the best is given by:
\begin{equation}
\hat{H} = \sum_{i=1}^{\hat{N}}\hat{w}_i\hat{H}_i,
\end{equation}
where $\hat{N} \leq n$.

The Hamiltonian obtained in this way acts as the ``discriminator'' and it used to train the generator in the typical minmax game of GANs.

\subsubsection{Generator}

The generator is a quantum computer that we use to prepare a quantum state
and evaluate the expectation values.
The generator is defined as a sum of parametrized quantum circuits
with associated probabilities (which can be interpreted as a quantum channel):
\begin{equation}
G(\theta) = \sum_{i=1}^r p_iG_i(\theta_i)\rho_0G_i(\theta_i)^\dagger.
\end{equation}
Here $\sum_{i=1}^r p_i = 1$ and $p_i$ is the probability associated with the
circuit $G_i$ and $\rho_0$ is the initial state of the circuit. The summation restricts the maximal rank of the output state that the generator is able to generate.
Namely, a generator of this form is able to learn a mix of at most $r$ pure states.

In all of our experiments we use the same design for each circuit within the
generator, i.e., $G_i = \bar{G} \quad \forall i \in 1,\ldots, r$, and only the
parameters $\theta_i$ differ. We use different designs for $\bar{G}$ for
different experiments. These fall into the following two categories:
\begin{enumerate}
\item Generic $\bar{G}$ described in Appendix~\ref{apx:sqgans_ansatz}.
\item $\bar{G}$ the same as the circuit used to generate the target state.
\end{enumerate}
We set the initial state to $\rho_0 = \ket{0}$ throughout.

In the end the generator becomes:
\begin{equation}
G(\theta) = \sum_{i=1}^r p_i\bar{G}(\theta_i)\ket{0}\bra{0}\bar{G}(\theta_i)^\dagger.
\end{equation}

\subsection{Training}

Similarly to WGANs the discriminator part of the qWGANs computes the best approximation of the quantum Wasserstein distance and the generator tries to minimize it.
This can be again encoded as the minmax game.

Given the target state $\rho_r$, the generated state $G(\theta)$ and using trace properties, Eq.~\eqref{eq:quantum_wasserstein_parametrized} can be stated as:
\begin{equation}
\label{eq:quantum_wasserstein_separated_trace}
\begin{split}
W_1(\rho_r, G(\theta)) %
&= \max_{w} \Tr\bigg[(\rho_r - G(\theta))\sum_{i=1}^N w_iH_i\bigg] \\
&= \max_{w} \sum_{i=1}^Nw_i(\Tr[\rho_rH_i] - \Tr[G(\theta)H_i]).
\end{split}
\end{equation}
Adding the goal of the generator to minimize the quantum Wasserstein distance from the target state, the qWGAN optimization objective reads:
\begin{multline}
\max_{w}{\min_{\theta}{\mathcal{L}(w, \theta)}} \\
= \max_{w} \min_{\theta} \sum_{i=1}^N w_i (\Tr[\rho_rH_i] - \Tr[G(\theta)H_i]).
\label{eq:qwgans_optimization_objective}
\end{multline}
In practice, the expectation $\Tr[\rho_r H_i]$ can be precomputed and re-used for the training process.

The overall training procedure is summarized in Algorithm~\ref{alg:qgan_learning}.

\begin{algorithm}[H]
\caption{WQGAN Learning}
\label{alg:qgan_learning}
\begin{algorithmic}[1] 
    \Require $\mathcal{H} = \{H_i\}$ - a set of Pauli strings to use for Wasserstein distance
    calculation
    \Require $\rho_r$ - a target state
    \State Compute a vector of expectations: \[ s = \left(\Tr[\rho_r H_1], \ldots, \Tr[\rho_r H_{\abs{\mathcal{H}}}]\right) \]
    \While{Stopping criterion}
        \State Compute \[ c_i = \Tr[G(\theta) H_i] - s_i \]
        \State Find $\hat{H}$ using the linear program from Eq.~\eqref{eq:linear_program_canonical}
        \State Use $\hat{H}$ to find the gradients of $\Tr[G(\theta)\hat{H}]$ w.r.t.\ the parameters $\theta_i$ and $p_i$ and update them
    \EndWhile
\end{algorithmic}
\end{algorithm}

\section{Extension to unseen state generation}
The shortcoming of the above formulation is the inability to generate
new, unseen states. In this chapter we propose the hybrid classical-quantum
method that extends qWGANs and allows to overcome this limitation.

Our idea is based on how the quantum Wasserstein distance is
approximated during the training. The discriminator at every step approximates
the distance between some fixed target state and the generated state which
changes after each iteration. However, the discriminator never needs an access to
the actual target state, it only operates on the set of measured
expectations.

Given a parametrized circuit $U$ and a set of parametrizations $\Theta =
\{\theta_i\}$ (where $\theta_i \in \mathbb{R}^l$ and $l$ is the number of
parameters in the circuit $U$) and a set of operators $H = \{H_j\}$, one can prepare the set of vectors of expectations $S$.
Each vector $s_{\theta_i} \in S$ contains the expectations of the circuit $U(\theta_i)$, such that $s_{\theta_i}^{(j)} = \langle H_j \rangle_{U(\theta_i)}$.

The proposed framework is defined in two parts as follows:
\begin{enumerate}
\item \textit{Classical}: Takes as the input the set $S$ and uses it to learn the function $f:
\mathbb{R}^{n} \to \mathbb{R}^{\abs{H}}$. Given a vector $g \in \mathbb{R}^n$, this function produces a new vector $s' = f(g)$ such that $\exists \theta' : (s')^{j} = \langle H_j \rangle_{U(\theta')}
\forall j$.
\item \textit{Quantum}: Takes $s'$ as the input and uses it as
the expectations of target state in the qWGANs setting described in the
previous chapter. The generator trained using $s'$ produces new, unseen before quantum state.
The qWGANs optimization objective from Eq.~\eqref{eq:qwgans_optimization_objective} becomes:
\begin{multline}
\max_{w}{\min_{\theta}{\mathcal{L}(w, \theta)}} \\
= \max_{w} \min_{\theta} \sum_{i=1}^N w_i \left(s^{\prime(i)} - \Tr[G(\theta)H_i]\right)
\label{eq:qwgans_optimization_objective_unseen}
\end{multline}
\end{enumerate}

Once the function $f$ is learned, it can be used arbitrary many times to produce
new vectors of expectations.
With those vectors, it is possible to generate new quantum states that come from
some circuit $U(\theta')$, without ever knowing $\theta'$ or even $U$.

\subsection{Labeled state generation}

If the quantum state produced by the circuit $U$ can be labeled by some continuous
variable, we can use this variable to find the function $f$.

Specifically, here we assume that each parameter $\theta_i^{(j)}$ is described by some
function, i.e., given a label $g_i \in V \subseteq \mathbb{R}$, $\forall \theta_i
\in \Theta$ $\theta_i = \theta(g_i) = [\theta^{(1)}(g_i), \theta^{(2)}(g_i),
\ldots, \theta^{(l)}(g_i)]$.
We also assume that the expectations of the state produced by the
circuit $U(\theta(g_i))$, can be described by some other continuous functions,
i.e., $\forall s_{\theta_i} \in S$ $s_{\theta_i} = s(g_i) = [s^{(1)}(g_i), s^{(2)}(g_i), \ldots,
s^{(\abs{H})}(g_i)]$, $s^{(j)}: V \to [-1; 1]\ \forall_{j \in 1,\ldots,\abs{H}}$.
Then, the input to the classical part of the framework is the set $S = \{s_{\theta_i}\}$, together with
corresponding set $G = \{g_i\}$.
To find $s^{(j)}\ \forall_{j=1,\ldots,\abs{H}}$ functions interpolation is
sufficient. So, the function $f: V \to \mathbb{R}^{\abs{H}}$ simply takes
any value of $g \in V$ and returns the expectations for this value using
interpolations of functions $s^{(j)}$.

Although the setup described here assumes a one-dimensional variable $g$, this
notion can be extended to a multi-variable case where $g \in V \subseteq
\mathbb{R}^m$.

\subsubsection{Application to Phase Transition}

This approach can be used when $U$ is the topological phase transition circuit
(Appendix~\ref{apx:topological_phase_transition_ansatz}) proposed by Smith et.~al \cite{smith2021}.
All the parameters of this circuit can be described by
three functions $\theta_v, \theta_w, \theta_r$ over $V = [-1; 1]$. To prepare
the input to the classical part $m$ ($m = \abs{S} = \abs{G}$) values of $g \in V$ are
sampled and the expectations of $U(\theta_v(g_i),
\theta_w(g_i), \theta_r(g_i))\ \forall_{i=1,\ldots,m}$ are calculated for all operators $H_i
\in H$. Similarly as in the WQGANs chapter, $H$ is chosen to be the set of all length-$k$ Pauli strings.
This data is used to interpolate the expectation functions for those operators.

\begin{figure}[!htb]
\centering
\includegraphics[width=1\linewidth]{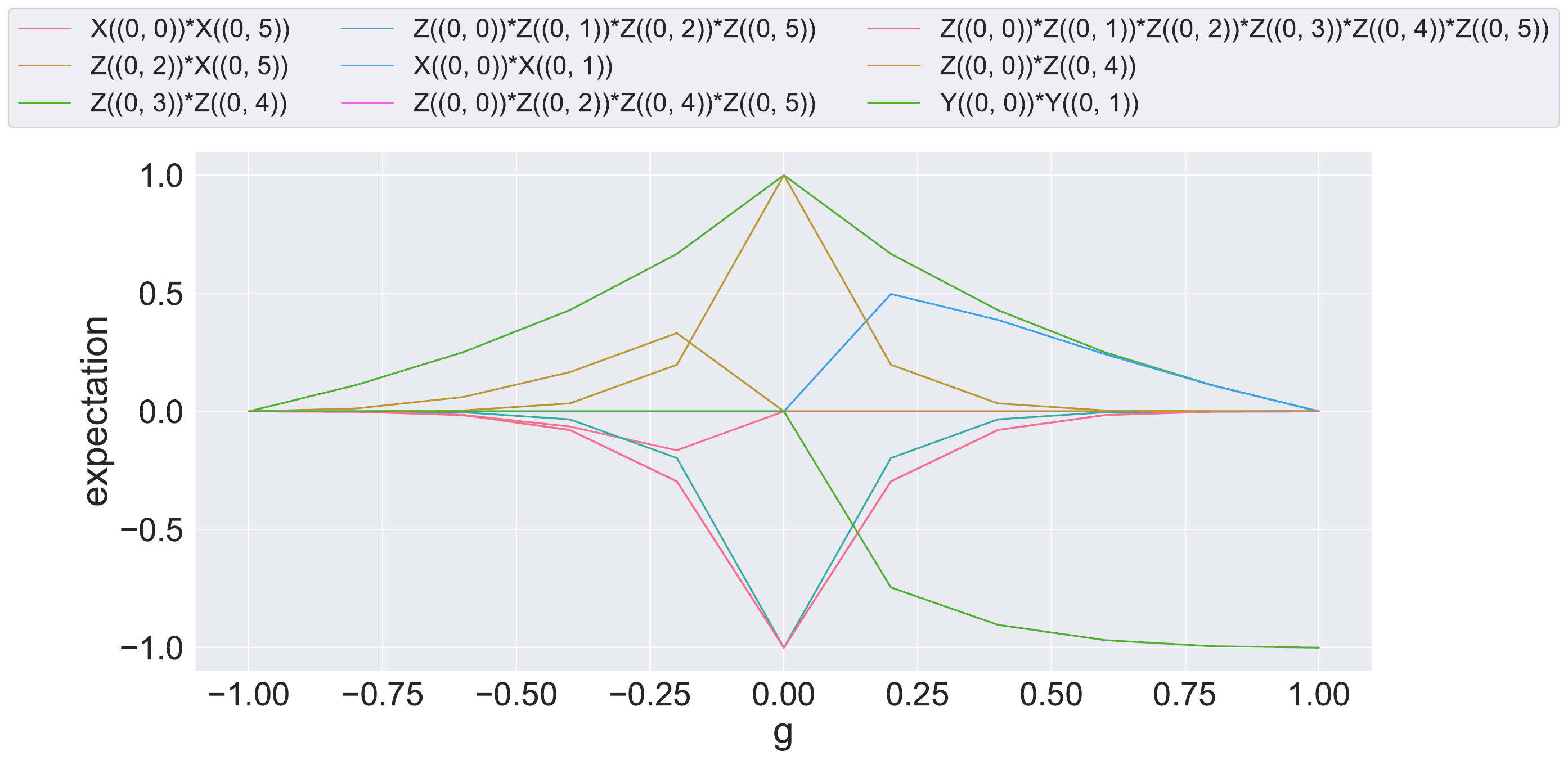}
\caption{Interpolated expectations of topological phase transition circuit (Appendix~\ref{apx:topological_phase_transition_ansatz}) with 5 qubits width for $9$ random 3-Pauli string operators.
Interpolation using evenly spaced 11 values of the parameter $g \in [-1; 1]$.}
\label{fig:phase_exps}
\end{figure}

In Fig.~\ref{fig:phase_exps} the interpolated expectations of the circuit for
$k=3$ and $m=11$ are plotted (only a subset of the expectation is plotted for readability).

\begin{figure}[!htb]
\captionsetup[subfigure]{labelformat=empty}
\centering
\subfloat{%
\includegraphics[width=0.5\linewidth]{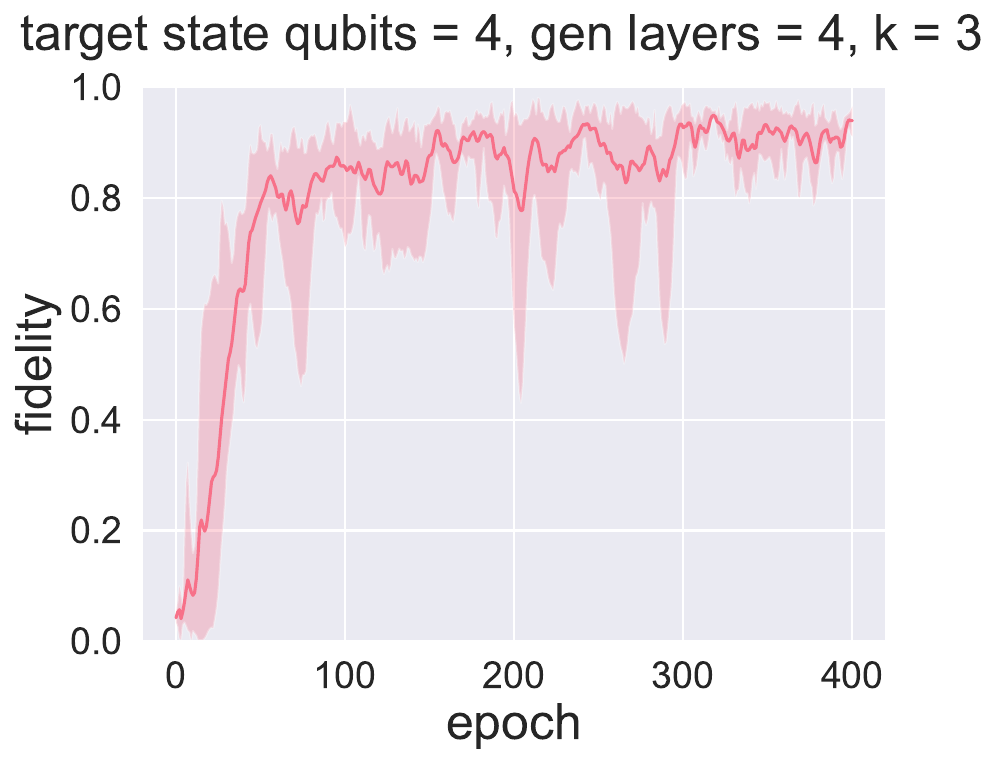}}
\subfloat{%
\includegraphics[width=0.5\linewidth]{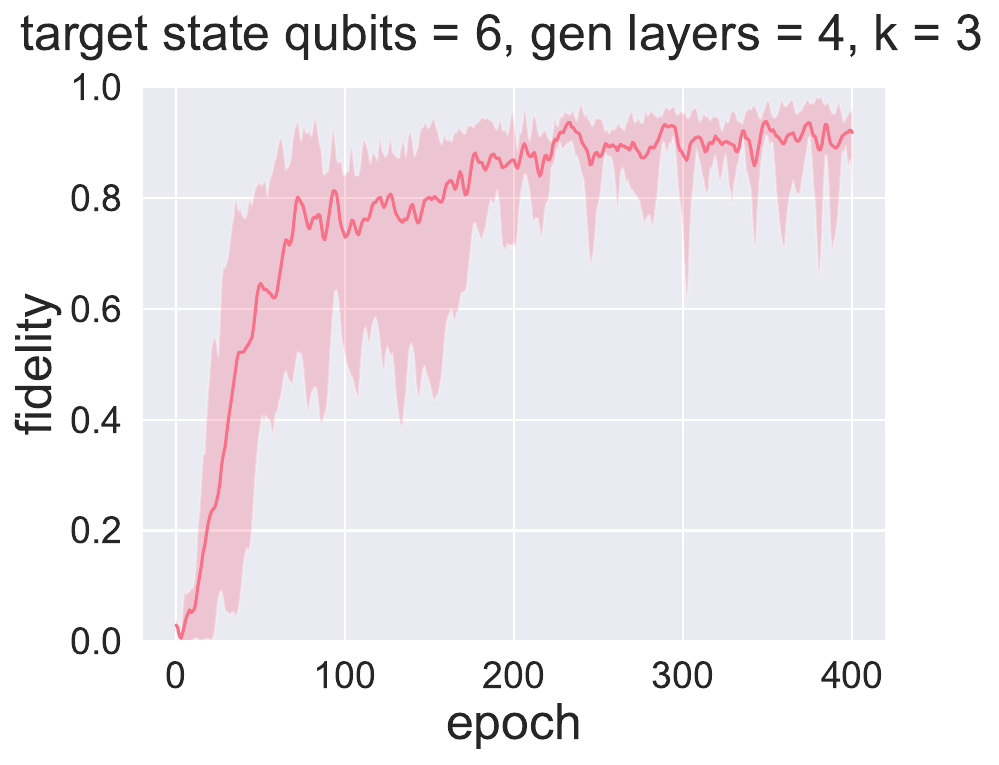}}

\subfloat{%
\includegraphics[width=0.5\linewidth]{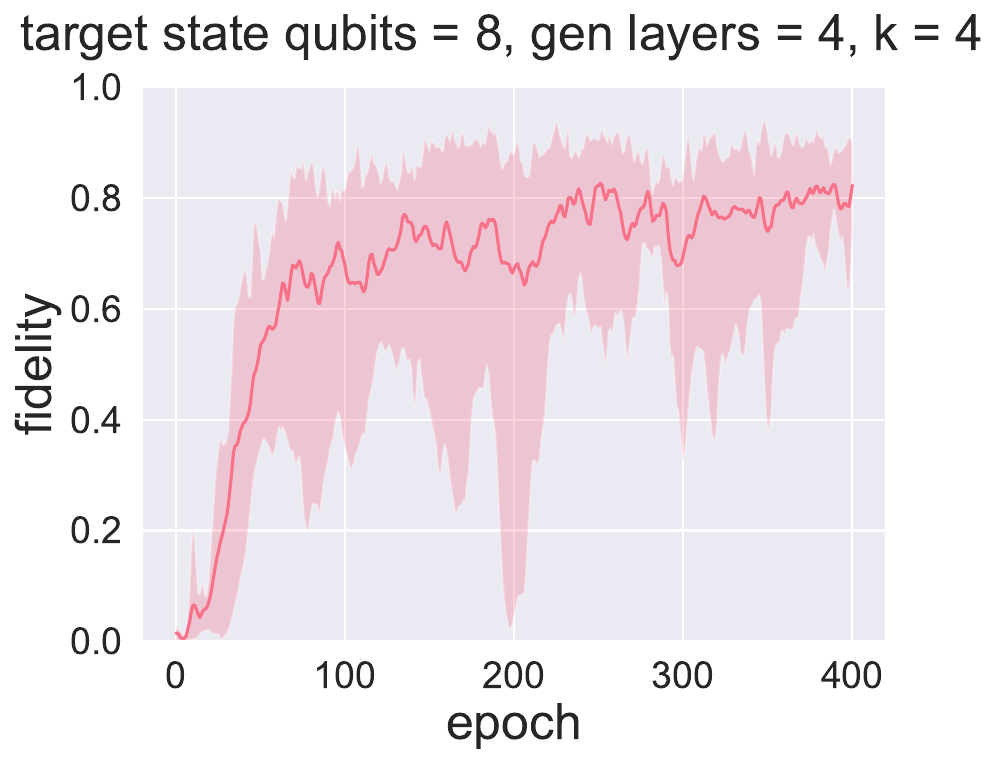}}
\subfloat{%
\includegraphics[width=0.5\linewidth]{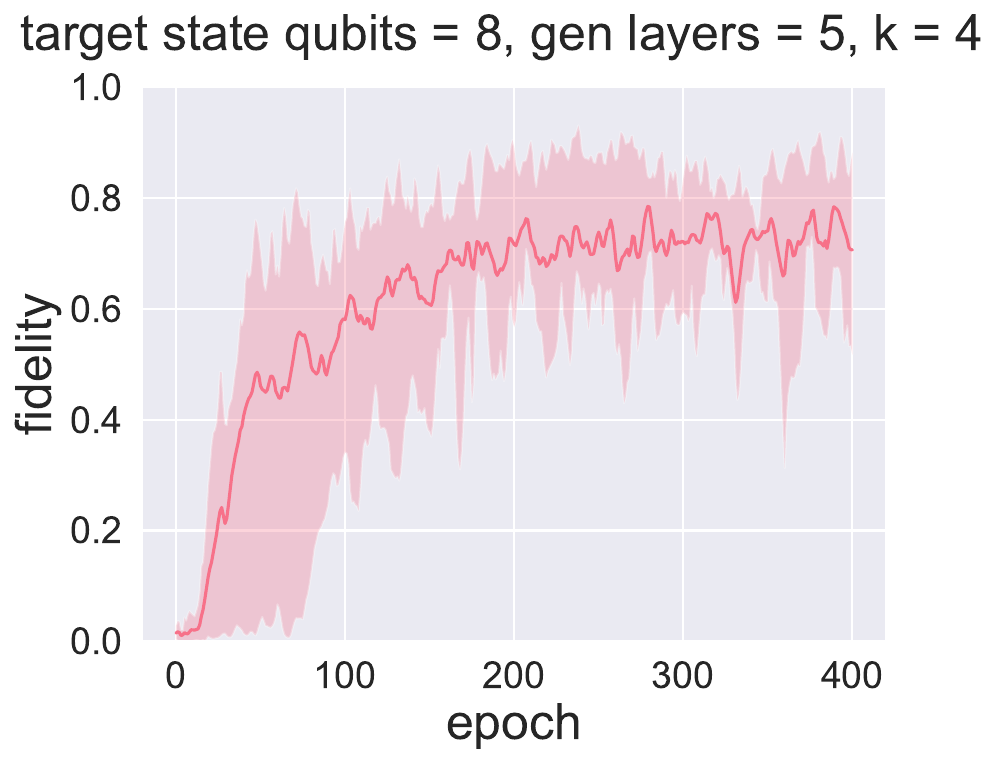}}

\subfloat{%
\includegraphics[width=0.5\linewidth]{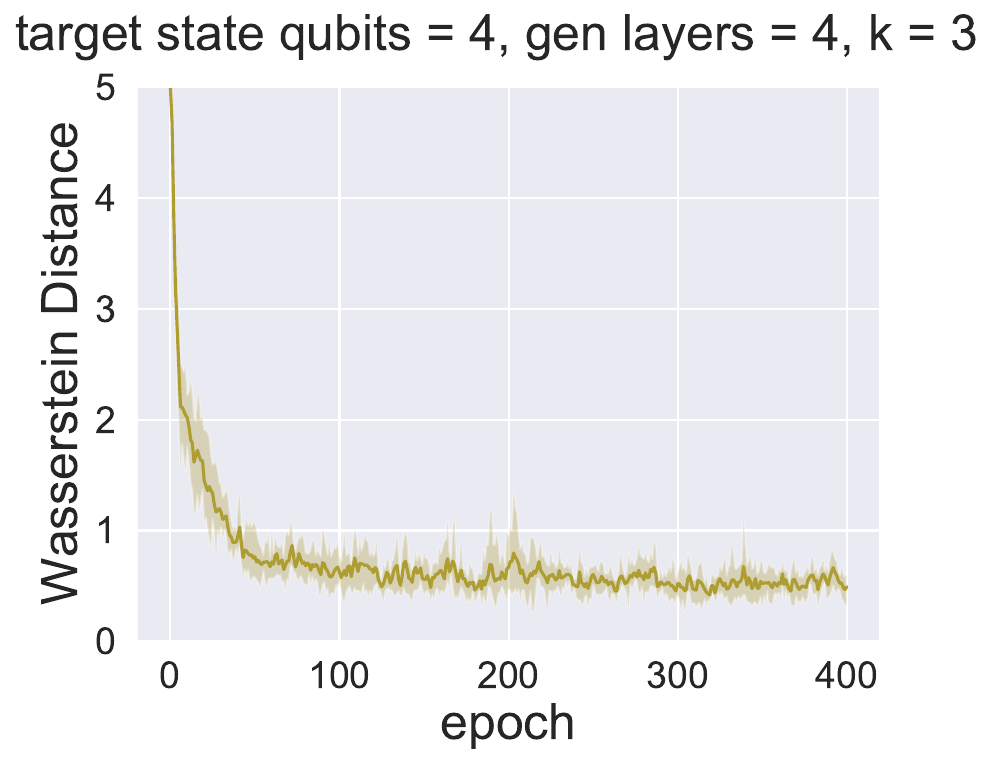}}
\subfloat{%
\includegraphics[width=0.5\linewidth]{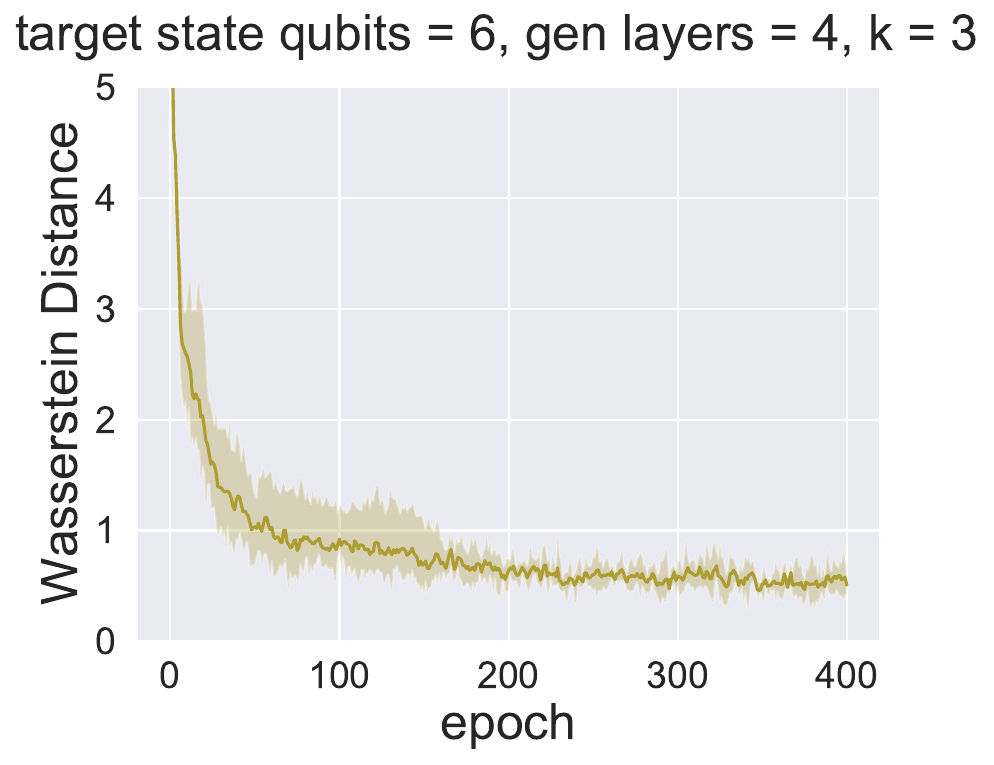}}

\subfloat{%
\includegraphics[width=0.5\linewidth]{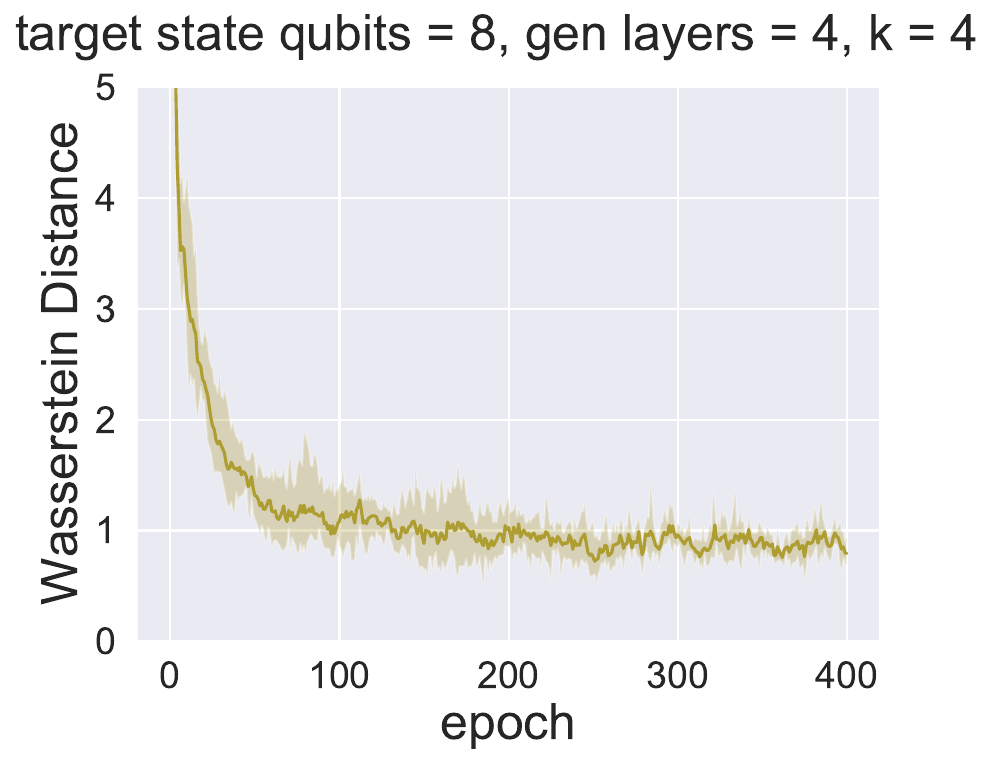}}
\subfloat{%
\includegraphics[width=0.5\linewidth]{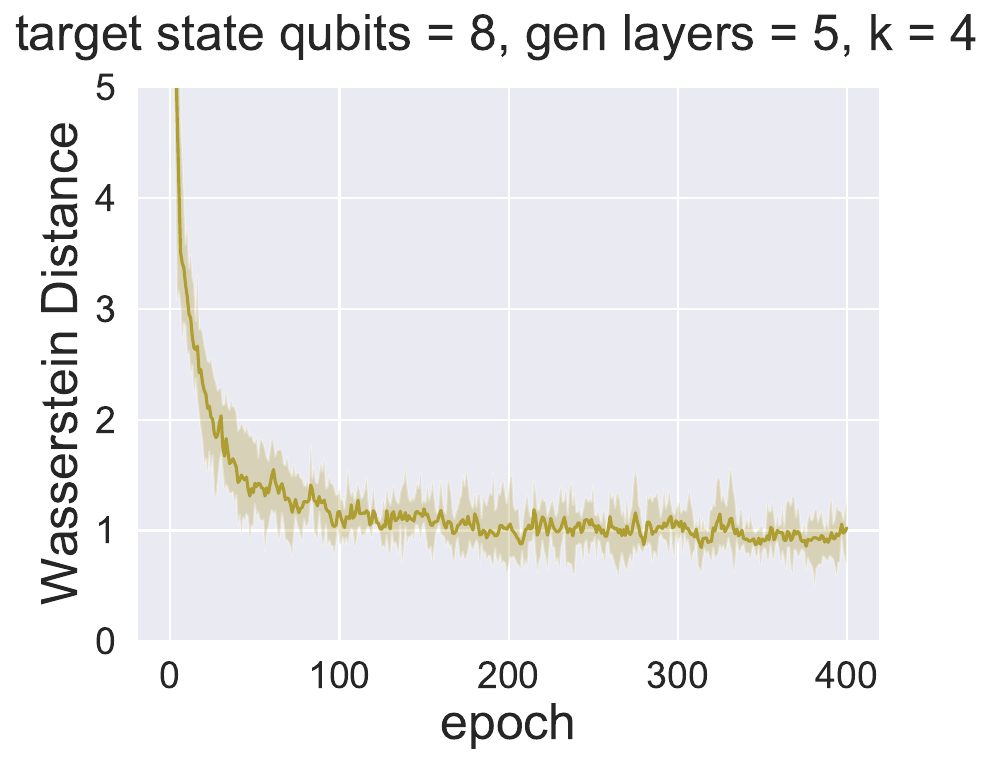}}
\caption{Results for the interpolated expectations of the topological phase transition circuit (Appendix~\ref{apx:topological_phase_transition_ansatz}) and the generator built 
with the generic circuit (Appendix~\ref{apx:sqgans_ansatz}).
The solid line represents the average value and the shaded area
represents the range from 5 different experiments. The upper row shows the
fidelity and the bottom row shows the corresponding Wasserstein distance.}
\label{fig:wqgans_res_interpolated_1}
\end{figure}

The interpolated expectations are used to learn the quantum states for
the values of $g$ that were not part of the classical input.
In Fig.~\ref{fig:wqgans_res_interpolated_1} we see the fidelity and Wasserstein distance
between the target states and the ones learned using the interpolated
expectations.

We can now use the states learned with the interpolated expectation to perform a
measurement of string order parameters and observe the phase transition.

\begin{figure*}[!htb]
\captionsetup[subfigure]{labelformat=empty}
\centering
\subfloat{%
\includegraphics[width=0.45\linewidth]{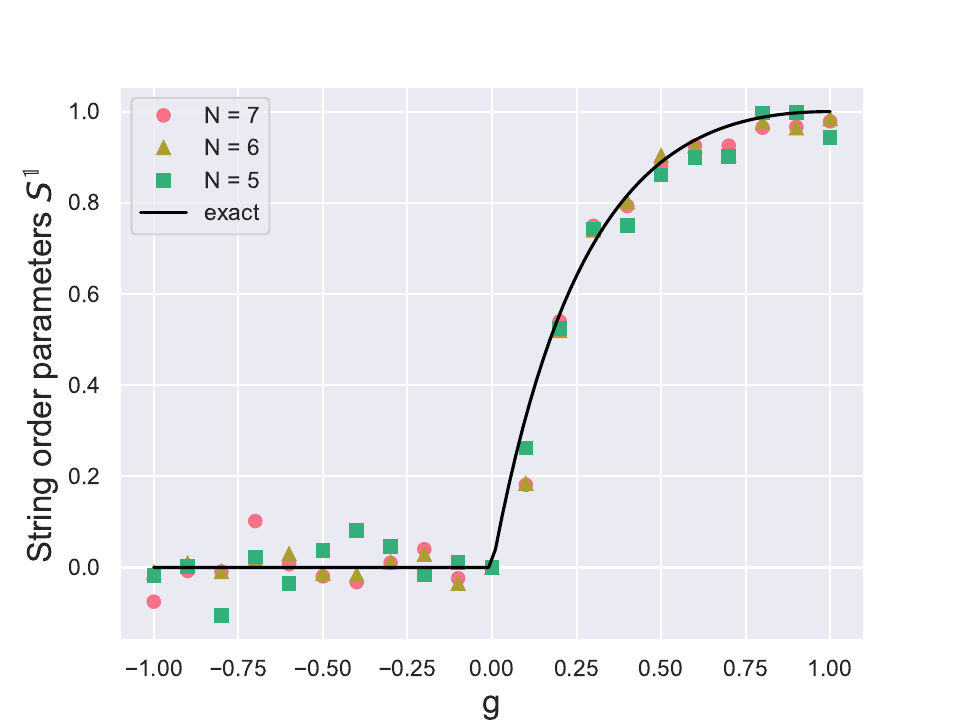}}
\hspace{0.05\linewidth}
\subfloat{%
\includegraphics[width=0.45\linewidth]{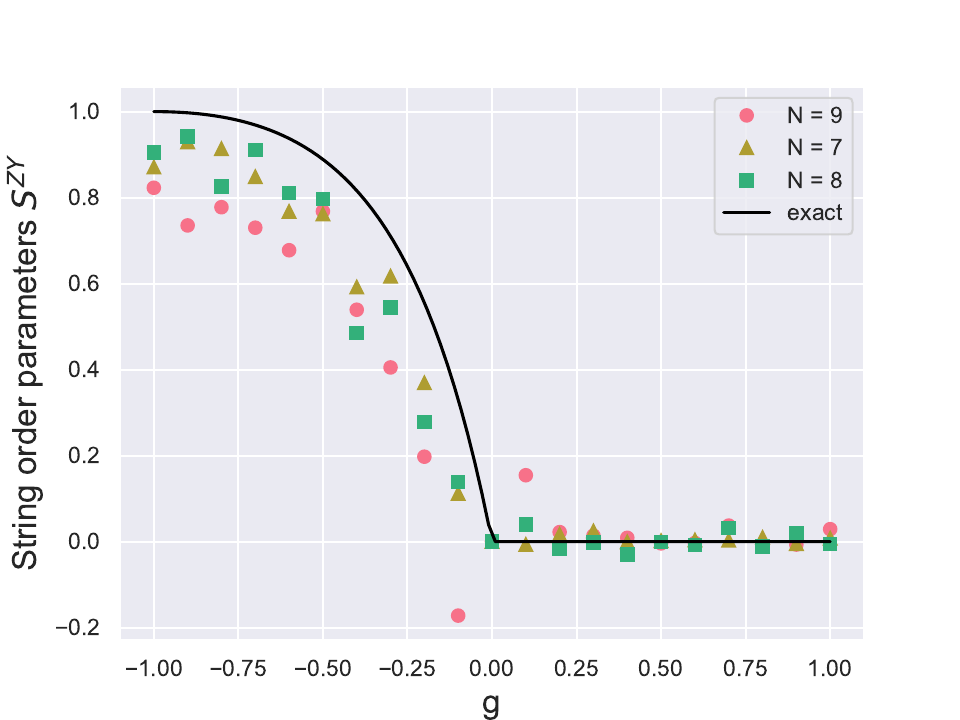}}

\caption{String order parameters $S^\mathbbm{1}$ and $S^{ZY}$ measured on the
generic generator from Appendix~\ref{apx:sqgans_ansatz}, trained using the
interpolated expectations, for different width of the circuit $N$.
The phase transition at $g=0$ is clearly visible,
the results are very close to the exact ones.}
\label{fig:string_order_1}
\end{figure*}

The string order parameters are defined as:
\begin{subequations}
\begin{align}
S^\mathbbm{1} &= \bra{\psi} \prod_{i=3}^{N-2} X_i \ket{\psi}, \\
S^{ZY}        &= \bra{\psi} Z_2 Y_3 \left(\prod_{i=4}^{N-3}X_i\right) Y_{N-2} Z_{N-1} \ket{\psi},
\end{align}
\end{subequations}
where $N$ is the width of the circuit and $\ket{\psi}$ is the final state
obtained by the topological phase transition circuit from Appendix~\ref{apx:topological_phase_transition_ansatz}.
The measurements of $S^{\mathbbm{1}}$ and $S^{ZY}$ on states learned using the
interpolated expectations are shown in Fig.~\ref{fig:string_order_1}. The
obtained results closely follow the expected value and the phase transition
point at $g=0$ is clearly distinguishable.

More importantly, in all experiments the generic generator ansatz from Appendix~\ref{apx:sqgans_ansatz} was used.
This means that the design of $U$ and its parametrization was unknown to the quantum generator and discriminator.

\subsection{Unlabeled state generation}

In more general case when the assumption about states being labeled does not hold,
other tools are need to find the function $f$. Here we make another
assumption, that all vectors in input set $S$ come from the same distribution
$p_S$. In such case, we can use the generative modeling to learn the function
$f$. In particular, here we use classical Wasserstein Generative Adversarial
Networks (WGANs) to approximate the distribution $p_S$ and later use the classical
generator as the function $f$ to produce the vectors $s'$.

We use this technique to generate new, previously unseen states from the butterfly circuit
(Appendix~\ref{apx:butterfly_ansatz}). First, we generate the set $S$ and use it
to train a simple WGAN-GP \cite{gulrajani2017}, with the penalty factor
$10$. We use simple 2-layers deep neural network (DNN), with input dimension $16$ and
with layer dimensions $64$ and $128$, for both, generator and
discriminator. We use Adam optimizer \cite{kingma2017adam} with the following
parameters: $\beta_1 = 0.9$, $\beta_2 = 0.999$, $\hat{\epsilon} = 1e - 7$ and the
learning rate of $0.001$. 

\begin{figure}[!htb]
\captionsetup[subfigure]{labelformat=empty}
\centering
\subfloat{%
\includegraphics[width=0.5\linewidth]{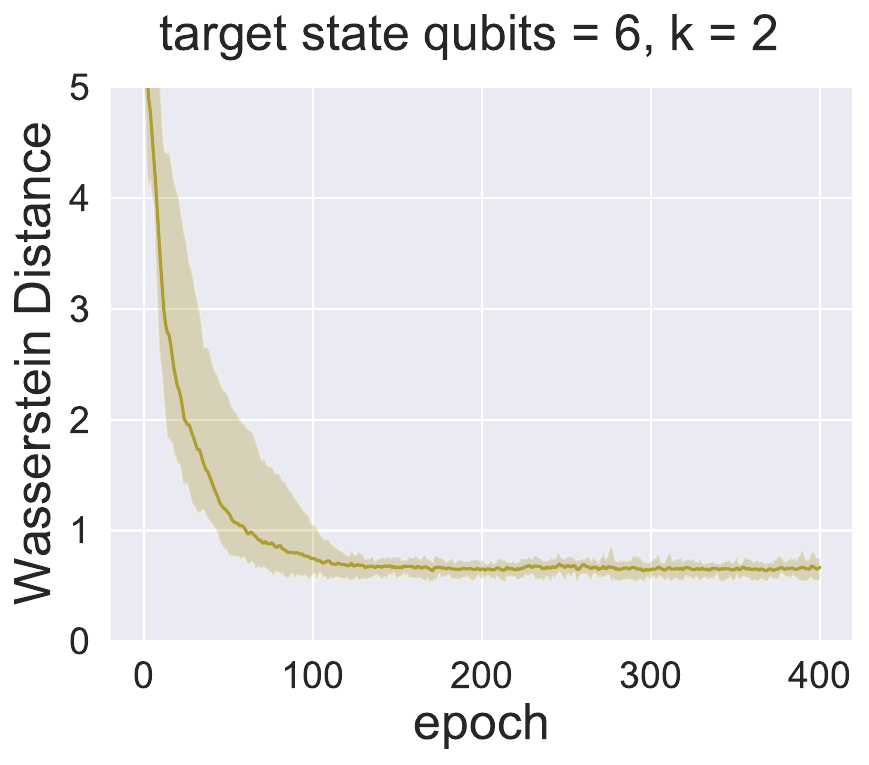}}
\subfloat{%
\includegraphics[width=0.5\linewidth]{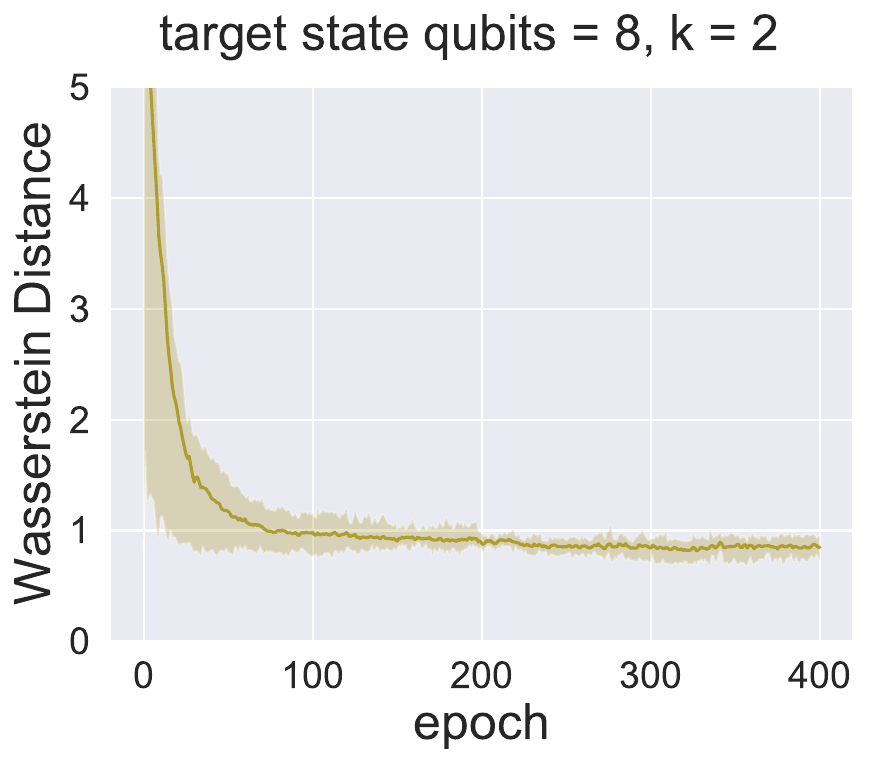}}
\caption{Wasserstein Distance for the expectations of the butterfly circuit
(Appendix~\ref{apx:butterfly_ansatz}) generated with GANs and the generator
build with the same butterfly circuit. Results
for training set size $\abs{S} = 256$ for $k = 4, 6$ and $\abs{S} = 512$ for $k = 8$.
The solid line represents the average value and the shaded area
represents the range from 5 different experiments.
}
\label{fig:wqgans_res_gans_1}
\end{figure}

For the states generated in this way, it is not possible to calculate the fidelity, so we
relay on the Wasserstein distance to evaluate the results. As shown in the previous
chapter, the decrease in the Wasserstein distance is strongly correlated with
the increase in fidelity. In Fig.~\ref{fig:wqgans_res_gans_1} results for
several different sizes of the target states are presented. We use the generator with the
same architecture as the target circuit, to see whether the expectations
generated by the classical GANs could be measured from the target circuit.
The Wasserstein distance very quickly drops below $1$, which should correspond to
the fidelity of more than $0.8$ based on the previous observations. However, it
always plateaus before dropping to 0, which indicates that the classical
generator does not produce expectations exactly from the $p_S$ distribution.

We have demonstrated the ability to generate unseen quantum state, with the
expectations generated by classical GANs. Despite using basic and shallow DNN
for the classical generator and discriminator, the generated expectations were
very close to the ones measured from the generated quantum state as indicated by
the measured Wasserstein distance. Using more sophisticated or deeper architecture
for the classical GANs could yield even better results or decrease the required training
size and is an interesting direction for further research.

\section{Conclusion}

The field of quantum machine learning is currently in an early, exploratory
stage. There have been many attempts to bring the successful classical machine
learning ideas into the quantum realm. In this work we took a closer look at
realization of Generative Adversarial Networks on quantum machines.

By leveraging WQGANs \cite{depalma2021} we proposed a new method
to generate unseen quantum states. We combined the classical generative modeling
with WQGANs to train a parametrized quantum circuit able to generate the unseen quantum
states. We showed in the numerical experiments that the quantum states generated with our
method can approximate the characteristic of the original source with high fidelity.

All the attempts so far, including this work, concentrated on small input of
several or maximum dozen qubits. An important are to explore is the
scalability of quantum GANs for wider inputs, especially on the currently available NISQ machines.
Another interesting question is, how could the unseen states be generated
in a purely quantum manner, without the need to use the classical computer.

\begin{acknowledgments}
This research is part of the Munich Quantum Valley, which is supported by the Bavarian state government with funds from the Hightech Agenda Bayern Plus.
\end{acknowledgments}

\appendix
\onecolumn
\section{Appendix}

\subsection{Circuits}

\subsubsection{Generic ansatz}
\label{apx:sqgans_ansatz}

\begin{figure*}[!htb]
\centering
\begin{tikzcd}[scale cd=0.9, font=\footnotesize]
    \qw &  \gate{R_x(\theta_{(x,1)}^{(i)})} & \gate{R_z(\theta_{(z,1)}^{(i)})} &
    \gate[2, disable auto height]{R_{zz}(\theta_{(1,2)}^{(i)})} & \qw & \qw \\
    \qw &  \gate{R_x(\theta_{(x,2)}^{(i)})} & \gate{R_z(\theta_{(z,2)}^{(i)})} &
    \qw & \gate[2, disable auto height]{R_{zz}(\theta_{(2,3)}^{(i)})} & \qw \\
    \qw &  \gate{R_x(\theta_{(x,3)}^{(i)})} & \gate{R_z(\theta_{(z,3)}^{(i)})} &
    \gate[2, disable auto height]{R_{zz}(\theta_{(3,4)}^{(i)})} & \qw  & \qw \\ 
    \qw &  \gate{R_x(\theta_{(x,4)}^{(i)})} & \gate{R_z(\theta_{(z,4)}^{(i)})} &
    \qw & \vdots \\
     & \vdots & \vdots & \vdots & \vdots & \\
    \qw &  \gate{R_x(\theta_{(x,w-2)}^{(i)})} & \gate{R_z(\theta_{(z,w-2)}^{(i)})} &
    \gate[2, disable auto height]{R_{zz}(\theta_{(w-2,w-1)}^{(i)})} & \vdots \\ 
    \qw &  \gate{R_x(\theta_{(x,w-1)}^{(i)})} & \gate{R_z(\theta_{(z,w-2)}^{(i)})} &
    \vdots & \gate[2, disable auto height]{R_{zz}(\theta_{(w-1,w)}^{(i)})} & \qw \\
    \qw & \gate{R_x(\theta_{(x,w)}^{(i)})} & \gate{R_z(\theta_{(z,w)}^{(i)})} &
    \qw & \qw  & \qw
\end{tikzcd}
\caption{Single layer of the generic ansatz used for generator circuits \cite{dallaire2018}. The vector $\theta$ contains the circuit parameters, with index $i$ denoting the layer number and $w$ the qubit wire. The layer can be repeated arbitrary many times.}
\label{fig:generator_circuit_apx}
\end{figure*}
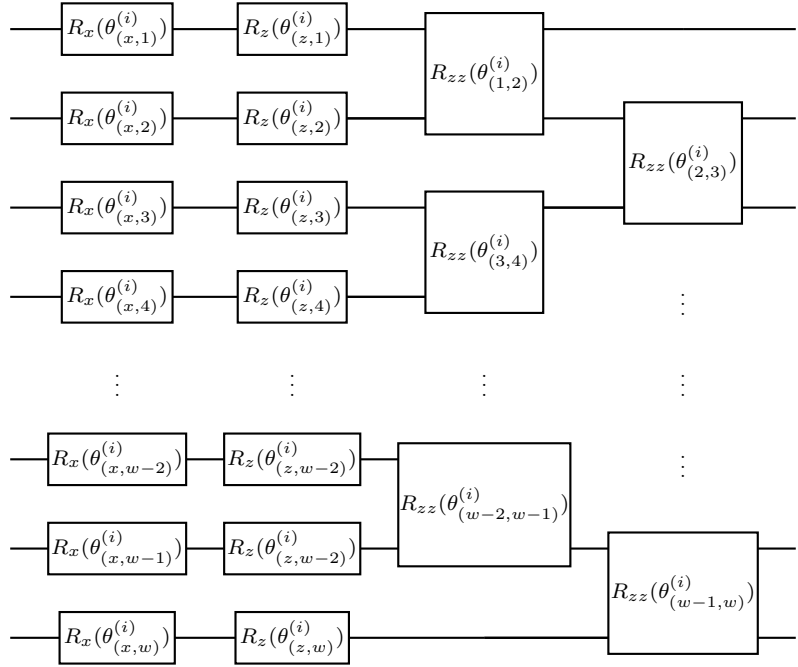

Fig.~\ref{fig:generator_circuit_apx} shows the generic circuit Ansatz used for the generator.

\FloatBarrier

\subsubsection{Topological Phase Transition ansatz}
\label{apx:topological_phase_transition_ansatz}

This circuit was used by Smith et al.~\cite{smith2021} to study
transitions between different states of matter. It is essentially a matrix
product state (MPS) represented in quantum circuit form.

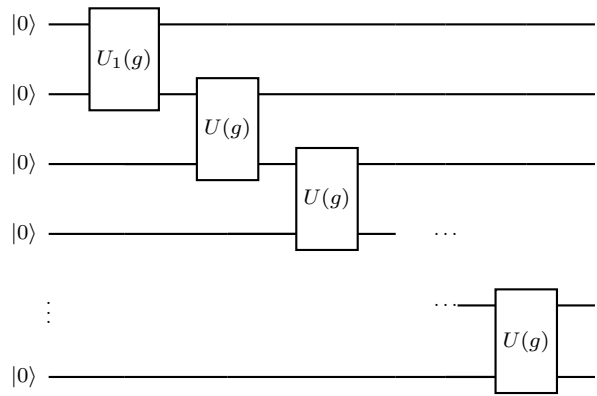
\begin{figure*}[!htb]
\centering
  \begin{tikzcd}[scale cd=0.9, font=\footnotesize]
    \lstick{$\ket{0}$} & \gate[2, disable auto height]{U_1(g)} & \qw & \qw & \qw &
    \qw & \qw & \qw \\
    \lstick{$\ket{0}$} & & \gate[2, disable auto height]{U(g)}  & \qw & \qw & \qw & \qw & \qw \\
    \lstick{$\ket{0}$} & \qw & \qw & \gate[2, disable auto height]{U(g)}  & \qw & \qw & \qw & \qw \\
    \lstick{$\ket{0}$} & \qw & \qw & \qw & \qw & \ldots  \\
    \vdots & & & & &\ldots & \gate[2, disable auto height]{U(g)} & \qw \\
    \lstick{$\ket{0}$} & \qw & \qw & \qw & \qw & \qw & \qw & \qw \\
  \end{tikzcd}
  \caption{The topological phase transition circuit studied in \cite{smith2021} }
  \label{fig:phase_circuit_apx}
\end{figure*}

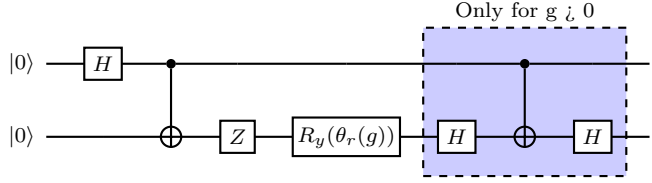
\begin{figure*}[!htb]
\centering
  \begin{tikzcd}[scale cd=0.9, font=\footnotesize]
    \lstick{$\ket{0}$} & \gate{H} & \ctrl{1} & \qw & \qw & \gategroup[2, steps=3,
    style={dashed, fill=blue!20, inner xsep=2pt}, background]{{
        Only for g > 0}}\qw & \ctrl{1} & \qw & \qw \\
    \lstick{$\ket{0}$} & \qw & \targ{} & \gate{Z} & \gate{R_y(\theta_r(g))} & \gate{H} & \targ{} &
    \gate{H} & \qw
  \end{tikzcd}
  \caption{The schema of $U_1$ gate from the circuit in Fig.~\ref{fig:phase_circuit_apx}}
  \label{fig:phase_circuit_u1}
\end{figure*}

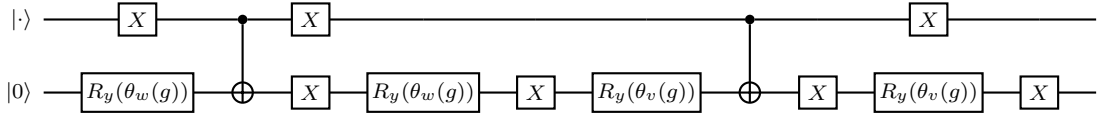
\begin{figure*}[!htb]
\centering
  \begin{tikzcd}[scale cd=0.9, font=\footnotesize]
    \lstick{$\ket{\cdot}$} & \gate{X} & \ctrl{1} & \gate{X} & \qw & \qw & \qw & \ctrl{1} &
     \qw & \gate{X} & \qw & \qw \\
     \lstick{$\ket{0}$} &  \gate{R_y(\theta_w(g))}  & \targ{} &  \gate{X} & \gate{R_y(\theta_w(g))} & \gate{X} &  \gate{R_y(\theta_v(g))}
     & \targ{} & \gate{X} & \gate{R_y(\theta_v(g))} & \gate{X} & \qw
  \end{tikzcd}
  \caption{The schema of $U$ gate from the circuit in Fig.~\ref{fig:phase_circuit_apx}}
  \label{fig:phase_circuit_u}
\end{figure*}
\FloatBarrier

Where the gates and parameters are defined as follows:
\begin{equation*}
\begin{split}
R_y(\theta) &= \begin{pmatrix}
  \cos{\frac{\theta}{2}} & -\sin{\frac{\theta}{2}} \\
  \sin{\frac{\theta}{2}} & \cos{\frac{\theta}{2}} 
\end{pmatrix}, \\
\theta_w(g) &= \arccos{\frac{\mathrm{sign}(g)\sqrt{\abs{g}}}{\sqrt{1+\abs{g}}}},\ \theta_w \in [0, \pi], \\
\theta_v(g) &= \arcsin{\frac{\sqrt{\abs{g}}}{\sqrt{1+\abs{g}}}},\ \theta_v \in [-\frac{\pi}{2}, \frac{\pi}{2}], \\
\theta_r(g) &= 2\arcsin{\frac{1}{\sqrt{1+\abs{g}}}},\ \theta_r \in [-\pi, \pi].
\end{split}
\end{equation*}

\subsubsection{Butterfly ansatz}
\label{apx:butterfly_ansatz}

\begin{figure*}[!htb]
\centering
  \begin{tikzcd}[scale cd=0.9, font=\footnotesize]
\lstick{$\ket{0}$} & \gate{R_x(\theta_{(1,1)})} & \ctrl{1} &
\gate{R_x(\theta_{(3,1)})} & \ctrl{2} & \qw & \gate{R_x(\theta_{(5,1)})} & \qw &
\ldots & \rstick[wires=9]{$\ket{\psi}$}\\
\lstick{$\ket{0}$} & \gate{R_x(\theta_{(1,2)})} & \gate{R_x(\theta_{(2,1)})}  &
\gate{R_x(\theta_{(3,2)})} & \qw & \ctrl{2} & \gate{R_x(\theta_{(5,2)})} & \qw &
\ldots & \\ 
\lstick{$\ket{0}$} & \gate{R_x(\theta_{(1,3)})} & \ctrl{1} &
\gate{R_x(\theta_{(3,3)})} & \gate{R_x(\theta{(4,1)})} & \qw & \gate{R_x(\theta_{(5,3)})} & \qw &
\ldots & \\
\lstick{$\ket{0}$} & \gate{R_x(\theta_{(1,4)})} & \gate{R_x(\theta_{(2,2)})}  &
\gate{R_x(\theta_{(3,4)})} & \qw & \gate{R_x(\theta{(4,2)})} & \gate{R_x(\theta_{(5,4)})} & \qw &
\ldots & \\ 
\lstick{$\ket{0}$} & \gate{R_x(\theta_{(1,5)})} & \ctrl{1} &
\gate{R_x(\theta_{(3,5)})} & \ctrl{2} & \qw & \gate{R_x(\theta_{(5,5)})} & \qw &
\ldots & \\ 
\lstick{$\ket{0}$} & \gate{R_x(\theta_{(1,6)})} & \gate{R_x(\theta_{(2,3)})}  &
\gate{R_x(\theta_{(3,6)})} & \qw & \ctrl{2} & \gate{R_x(\theta_{(5,6)})} & \qw &
\ldots & \\ 
\lstick{$\ket{0}$} & \gate{R_x(\theta_{(1,7)})} & \ctrl{1} &
\gate{R_x(\theta_{(3,7)})} & \gate{R_x(\theta{(4,3)})} & \qw & \gate{R_x(\theta_{(5,7)})} & \qw &
\ldots & \\ 
\lstick{$\ket{0}$} & \gate{R_x(\theta_{(1,8)})} & \gate{R_x(\theta_{(2,4)})}  &
\gate{R_x(\theta_{(3,8)})} & \qw & \gate{R_x(\theta{(4,4)})} & \gate{R_x(\theta_{(5,8)})} & \qw &
\ldots & \\
\lstick{$\ket{0}$} & \gate{R_x(\theta_{(1,9)})} & \qw  &
\gate{R_x(\theta_{(3,9)})} & \qw & \qw & \gate{R_x(\theta_{(5,9)})} & \qw &
\ldots & \\
\end{tikzcd}
\caption{The butterfly circuit for 9 qubits. For each $j$-th power of $2$ that the
  width of the circuit exceeds, the next layer is added that consist of $R_x$
  gates on each qubit and controlled $R_x$ gate between $i$-th and $i+2^j$-th qubits (continued below).}
\end{figure*}
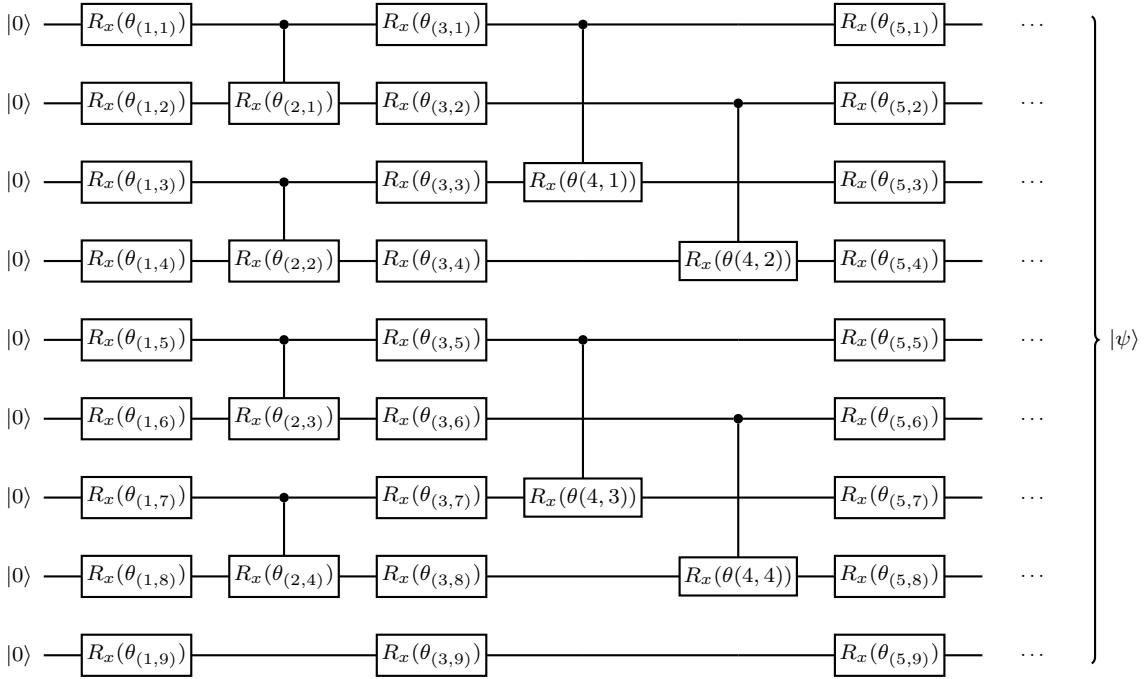

\begin{figure*}[htbp!] \ContinuedFloat
\centering
  \begin{tikzcd}[scale cd=0.9, font=\footnotesize]
  \lstick[wires=9]{$\psi$} & \ldots & \ctrl{4} & \qw & \qw &
  \qw & \gate{R_x(\theta_{(7,1)})} & \ctrl{8} & \qw \\
  & \ldots & \qw & \ctrl{4} & \qw & \qw  &
  \gate{R_x(\theta_{(7,2)})} & \qw & \qw\\
  & \ldots & \qw & \qw & \ctrl{4} & \qw &
  \gate{R_x(\theta_{(7,3)})}  \qw & \qw & \qw\\
  & \ldots & \qw & \qw & \qw & \ctrl{4}  &
  \gate{R_x(\theta_{(7,4)})}  \qw & \qw & \qw\\
  &  \ldots &\gate{R_x(\theta_{(6,1)})} & \qw & \qw & \qw &
  \gate{R_x(\theta_{(7,5)})}  \qw & \qw & \qw\\
  & \ldots & \qw & \gate{R_x(\theta_{(6,2)})}  & \qw & \qw &
  \gate{R_x(\theta_{(7,6)})}  \qw & \qw & \qw\\
  & \ldots & \qw & \qw & \gate{R_x(\theta_{(6,3)})} & \qw &
  \gate{R_x(\theta_{(7,7)})}  \qw & \qw & \qw\\ 
  &  \ldots &\qw & \qw & \qw & \gate{R_x(\theta_{(6,4)})}  &
  \gate{R_x(\theta_{(7,8)})}  \qw & \qw & \qw\\
  & \ldots & \qw & \qw & \qw & \qw &
  \gate{R_x(\theta_{(7,9)})}  & \gate{R_x(\theta_{(8,1)})} & \qw \\
\end{tikzcd}
\caption{The butterfly circuit for 9 qubits. For each $j$-th power of $2$ that the
    width of the circuit exceeds, the next layer is added that consist of $R_x$
    gates on each qubit and controlled $R_x$ gate between $i$-th and $i+2^j$-th qubits }
\end{figure*}

\FloatBarrier
\twocolumn

\bibliographystyle{quantum}
\bibliography{references}

\end{document}